\documentclass[twocolumn,showpacs,preprintnumbers,amsmath,amssymb,aps,prl]{revtex4}
\usepackage{graphicx}
\begin{document}

\title{Pattern Switching and Polarizability For Colloids in Optical Trap Arrays}   
\author{C. Reichhardt and    
 C. J. Olson Reichhardt} 
\affiliation{
Theoretical Division,
Los Alamos National Laboratory, Los Alamos, New Mexico 87545} 

\date{\today}
\begin{abstract}
We show that colloidal molecular crystal states 
interacting with a periodic substrate, such as an optical trap array,   
and a rotating external field
can undergo a rapid pattern switching in which the orientation of the 
crystal changes.
In some cases, a martensitic-like symmetry switching occurs.
It is also possible to create a 
polarized state where the colloids in each substrate minima develop
a director field which smoothly rotates with the external drive, 
similar to liquid crystal behavior.
These results open the possibility for creating
novel types of devices using photonic band gap materials, and 
should be generalizable to a variety of other condensed matter systems
with multiple particle trapping.   
\end{abstract}
\maketitle

\vskip2pc
There are a wealth of systems with commensurate and incommensurate states
which 
can be modeled as interacting particles on 
ordered substrates,
including vortices in type-II superconductors with periodic 
pinning assemblies \cite{Baert}, 
vortices in Bose-Einstein condensates with optical trap arrays 
\cite{Peeters}, 
atoms on surfaces \cite{Atom},  
electrons on periodic landscapes \cite{Peeters2},
and cold atoms or molecules in optical traps 
where Wigner crystal states can occur \cite{Wu}. 
Another system of this type that has been attracting growing interest 
is charge-stabilized colloidal particles in the 
presence of optical trap arrays, where a remarkably wide
variety of crystalline states 
\cite{Reichhardt,Bechinger,Frey,Trizac,Bechinger2,Bech},
sliding dynamics \cite{Korda,Olson},
and nonequilibrium phenomena \cite{Ladavac} can be realized.
An attractive aspect of the colloidal system 
is that the dynamics of the individual particles can 
be observed directly, making colloids on periodic substrates an ideal system 
for studying new types of behaviors 
occurring on periodic substrates that could be realized on
much smaller scales.  
Additionally, the ability to create and rapidly 
control colloidal crystal structures has a wide range of applications
in photonic and phononic materials, optical switches \cite{Vos}, 
photonic band-gap materials \cite{Bech},
and self-assembly of nanostructures. 

Crystalline states known as colloidal molecular crystals (CMCs), 
which have an additional orientational degree of freedom, have been shown
to form in the presence of a two-dimensional periodic substrate when there
are an integer number of colloids per substrate minimum 
\cite{Reichhardt,Bechinger,Frey,Trizac}.
The orientational degree of freedom permits the
equilibrium colloidal states to exhibit
spin type ordering, including a ferromagnetic state where all the CMCs 
point in the same direction, an antiferromagnetic state, 
and herringbone states \cite{Frey,Trizac}. 
Here we show that when an in-plane external rotating electric 
field is applied with a magnitude small enough that the
colloids remain confined within the 
individual traps, the orientational ordering, structure, 
and direction of the CMCs can be directly controlled.
In particular, we find that 
ferromagnetically interacting CMCs can undergo a switching
phenomenon 
where the orientation rapidly switches between different directions. 
For certain regimes, the colloids in each substrate minima polarize and 
develop a
director field which smoothly follows
the rotating external drive, similar to a liquid crystal state.  
For dimers with herringbone ordering, the rotating drive can induce symmetry 
changes in the CMC structure, giving rise to a martensitic switching behavior.  
The ability to create colloidal crystal structures that can rapidly 
switch among different orientations and structures 
could have a profound impact in creating new types of devices utilizing
photonic band gap materials, such as sensors, optical switches, 
and filters. 
Similar rapid structure transitions have already been achieved by applying
mechanical stress to macroscopic elastic hole structures
\cite{Mullin,Boyce}.
Here we show that similar types of 
structural transitions occur for colloids
subjected to an external field rather than to mechanical stress.   
The general physics of the colloidal system studied in this work 
may also be realized for other systems of interacting particles 
on periodic substrates, such as ions or localized electrons 
on ordered substrates or cold-atom Wigner crystal states 
in the presence of periodic optical trap arrays. 
If similar polarization and switching dynamics can be 
realized in these systems, a
new class of nanodevices could be explored.  

We model an assembly of $N_c$ colloidal particles
interacting with a two-dimensional triangular optical trap array at 
fillings of either two or three colloids per trap. 
The system has periodic boundary conditions in the $x$ and $y$-directions 
and the dynamics of the colloids are obtained by 
integrating the following overdamped equation of motion:  
\begin{equation} 
\eta\frac{d{\bf R}_{i}}{dt} = \sum^{N_c}_{j = 1}- \nabla V(R_{ij}) +
{\bf F}_{s} + {\bf F}_{ext} .
\end{equation} 
The colloid-colloid interaction is repulsive and has the screened Coulomb form 
$V(R_{ij}) = (E_{0}/R_{ij})\exp(-\kappa R_{ij})$
where $E_{o} = Z^{*2}/4\pi \epsilon\epsilon_{0}a_{0}$, 
$Z^{*}$ is the unit of charge, $\epsilon$ is the solvent dielectric constant,
$\eta$ is the damping constant,
$R_{ij}=|{\bf R}_i-{\bf R}_j|$, 
and ${\bf R}_{i(j)}$ is the position of of colloid $i(j)$.      
The system size is measured in units of $a_{0}$ and  
the screening length $1/\kappa$ is fixed at $2a_{0}$. 
Force and time are measured in units of $\tau = \eta/E_{0}$ and
$F_{0} = E_{0}/a_{0}$. 
We neglect hydrodynamic interactions since we are working in the
low volume, highly charged, electrophoretic limit \cite{noHI}.
The substrate force comes from a
triangular substrate of strength $F_s$ with
${\bf F}_{s} = \sum^{3}_{k=1}F_s\sin(2\pi b_{k}/a_{0})[\cos(\theta_{k}){\hat x} -
\sin(\theta_{k}){\hat y}]$
where $b_{k} = x\cos(\theta_{k}) -y\sin(\theta_{k}) + a_{0}/2$,
$\theta_{1} = \pi/6$, $\theta_{2} = \pi/2$, and
$\theta_{3} = 5\pi/6$. 
The initial colloidal positions are found by simulated annealing
\cite{Reichhardt}, and
we obtain the same ordered ground states 
observed in previous studies \cite{Reichhardt,Bechinger,Trizac}.  
After annealing, a circular ac drive of amplitude $A_{ac}$ is applied with   
${\bf F}_{ext} = A_{ac}\sin(2\pi\nu t){\hat {\bf x}} + A_{ac}\cos(2\pi\nu t){\hat {\bf y}}$, where $\nu=5 \times 10^{-6}$ inverse simulation time steps.

\begin{figure}
\includegraphics[width=3.5in]{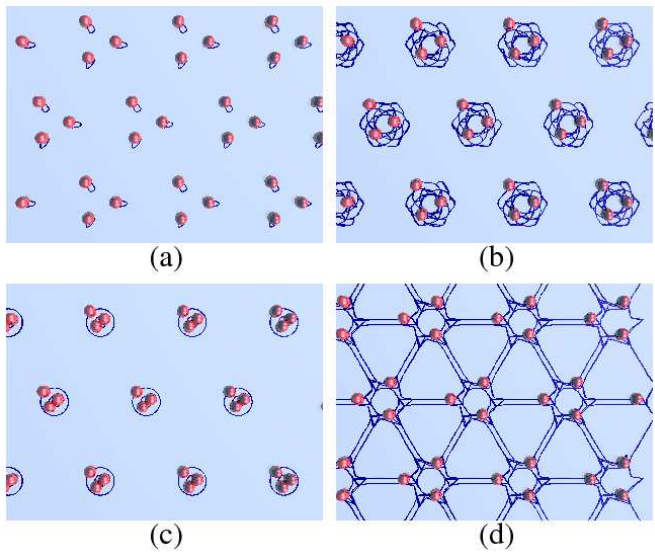}
\caption{
Colloid positions (balls) and trajectories (lines) 
in a portion of the sample over several ac drive
periods for a
system with 
three colloids per substrate minima. 
(a) In the locked (L) phase at $F_s  = 2.5$ and $A_{ac} = 0.5$, 
the trimers remain fixed in the same orientation. 
(b) In the switching (S) phase at $F_s = 4.0$  and $A_{ac} = 1.45$, 
the trimers  
switch between six different orientations in a single drive period.
(c) In the continuously polarized (CP) phase at $F_s = 6.5$ and $A_{ac} = 2.5$, 
the trimers rotate continuously with the ac drive.
(d) In the partially depinned phase at 
$F_s = 2.5$ and $A_{ac} = 0.65$, 
one of the possible dynamical modes is shown where
one third of the colloids depin and
switch in a large triangular pattern while the other colloids 
remain pinned and switch in 
a smaller star pattern \cite{M}.
}
\end{figure}

We first study the case of three colloids per potential minima.
The trimers that form in each trap are all aligned in the
same direction via an effective ferromagnetic coupling with neighboring
trimers
\cite{Reichhardt,Bechinger,Frey}. 
Under a rotating external drive, five distinct dynamical responses
appear.
In Fig.~1(a) we plot the trimer trajectories over several ac
drive periods for a system with $F_s = 2.5$ and $A_{ac} = 0.5$ in the locked (L)
phase, where 
the trimers undergo a small elliptical motion but remain aligned in a
fixed orientation 
\cite{M}.  The trajectories at 
$F_s = 4.0$ and $A_{ac} = 1.45$, 
shown in Fig.~1(b),
are much more complicated and the distorted circular orbits indicate 
that the trimers change their orientation periodically, 
leading us to term this the
switching (S) phase.
Here the triangle formed by the colloids in each
trimer is no longer equilateral; instead, one colloid moves further away from
the other two, elongating the trimer.
The resulting polarized trimer has a director, and in the lowest energy
state this director would remain aligned with the rotating drive; 
however, the outer colloid of the trimer
experiences a corrugated potential generated by the 
underlying substrate
and is unable to rotate smoothly.
Instead, the trimer first undergoes a slow continuous displacement and begins
to compress, followed by an abrupt switching from one symmetry direction
of the triangular substrate to another.
Within a single ac period there are six switching events, as shown 
in Fig.~2(b) 
where we present the instantaneous $x$-component $V_x$ of the
velocity for the system in Fig.~1(b). The smooth 
sinusoidal motion is superimposed with six sharp spikes per period. 
In Fig.~1(c) we show that at $F_s = 6.5$ and $A_{ac} = 2.5$ the 
trimers are more strongly polarized
and the director continuously follows the ac drive, 
as indicated by the two sets of circular trajectories. 
In Fig.~2(a) the corresponding $V_x$ 
has only a smooth sinusoidal motion. 
The ability of the director of the polarized trimer 
to couple with and follow an external field has many 
similarities to liquid crystal systems and
opens the possibility of realizing colloidal versions of 
liquid crystal devices. 

\begin{figure}
\includegraphics[width=3.5in]{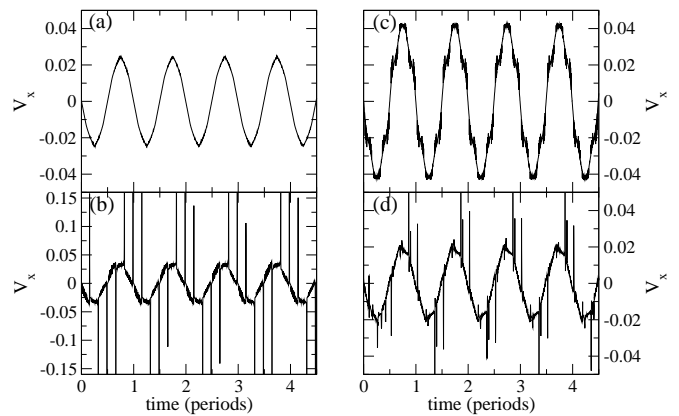}
\caption{
The instantaneous $x$-component $V_x$ of the colloid velocity.
(a) The trimer CP phase in Fig.~1(c) showing a smooth sinusoidal curve.
(b) The trimer S phase from Fig.~1(b) showing a sinusoidal curve
with six spikes per period that correspond to the switching events.
(c) The dimer CP regime for the system in Fig.~4(f) where the
dimers continuously rotate. 
(d) The dimer S phase for the system in Fig.~4(b). 
}
\end{figure}

Phases with multiple switching dynamics can also appear.  For example,
Fig.~1(d) illustrates a partially depinned state 
at $F_s=2.5$ and $A_{ac}=0.65$ in the
trimer system 
where one third of the colloids depin and move in large
triangular orbits while the other colloids remain pinned and
form smaller asymmetric 
star orbits. 
The velocities $V_x$ (not shown) in this regime have spikes similar
to those in Fig.~2(b). 
For $A_{ac} > F_s$, the system enters a new regime where the colloids 
continuously depin and
undergo sliding dynamics, as described in Ref.~\cite{Olson}.

Fig.~3(a) shows the phase diagram of the different regimes 
for $A_{ac}$ vs $F_s$. For weak $F_s$, the S phase appears only 
in a small window.
The CP phase occurs for $F_s>5.5$, and the width of this
phase increases with increasing $F_s$
since the trimers are increasingly 
compressed by the stronger trapping potential. 
Near the center of the trap, the potential appears
parabolic, so the substrate symmetry that determines the orientation of the 
trimers become less important and the discrete switching of the S phase
cannot occur for the compressed trimers.
A partially depinned phase such as that shown in Fig.~1(d), 
in which only a portion of the colloids are
depinned, occurs in a narrow region below the sliding or depinned phase.  

\begin{figure}
\includegraphics[width=3.5in]{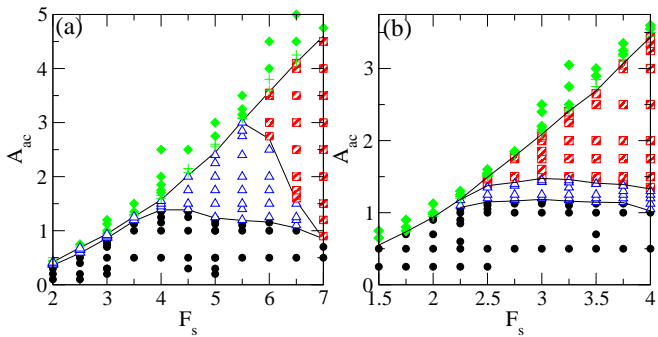}
\caption{
Phase diagrams as a function of ac amplitude $A_{ac}$ versus 
substrate strength $F_s$. Filled circles: locked (L) phase; open
triangles: switching (S) phase; hatched squares: continuously polarized (CP)
phase; plus signs: partially depinned phase; diamonds: depinned phase.
(a) Trimer state. The L phase is illustrated in Fig.~1(a), the S phase in
Fig.~1(b), the CP phase in Fig.~1(c), and the partially depinned phase
in Fig.~1(d).  (b) Dimer state.  The L HB phase is illustrated in Fig.~4(a),
the S HB phase in Fig.~4(b), and the CP phase in Fig.~4(f).  Along the
boundary of the S and CP phases we find the martensitic switching regime which
is a combination of S and CP.
}
\end{figure}

We next consider the case of two colloids, which form dimers, per 
substrate minimum.
The ground state has a herringbone (HB) structure, as in Fig.~4(e);
there are three possible orientations. 
In the locked (L) phase, illustrated at $F_s = 3.0$ and $A_{ac} = 1.0$
in Fig.~4(a), the colloids undergo small rotations but the orientation 
of the HB state remains fixed.
The switching (S) phase is more complicated for the 
dimer HB phase than for the trimers, and the complex orbits of
the dimers are illustrated in Fig.~4(b) for $F_s=3.0$ and $A_{ac}=1.25$.
Each switch occurs in three stages.  First, there is an abrupt buckling 
of every other HB row such that the dimers in the buckled row 
are alternately oriented perpendicular and parallel to the HB axis.  This
is illustrated in Fig.~4(c); the HB orientation prior to the switch
was horizontal as in Fig.~4(a).  After the sudden buckling, the dimers
in the buckled rows begin to tilt gradually, and at the same time the dimers
in the unbuckled rows also begin to tilt, as shown in Fig.~4(d).  Finally,
there is a second sudden switch when the dimers all snap into the new 
orientation, seen in Fig.~4(e).  This same pattern of switching occurs for
all three HB orientations.
The sharp switches appear as spikes in $V_x$ in Fig.~2(d).

For higher $A_{ac}$, the dimers lose the HB orientation and instead form the
fully polarized or ferromagnetic (FM) state illustrated in
Fig.~4(f) for $F_s = 3.0$ and $A_{ac} = 2.0$.    
In this continuously polarized (CP) regime, the dimers rotate 
continuously with the external drive, 
as 
indicated by the sinusoidal velocity response in Fig.~2(c).
Between the S and CP regimes, the system exhibits a combination 
of the S and CP phases where the switching HB configuration can 
change into the FM configuration and rotate continuously for a period of
time before jumping back to the HB switching state again. 
This behavior can be viewed as 
martensitic switching between two different
crystalline symmetries.  

In Fig.~3(b), the phase diagram for the dimer system as a function of $F_s$
and $A_{ac}$ indicates that the S and CP phases appear only for $F_s>2.25$.
The CP phase grows rapidly in extent with increasing $F_s$ due
to the reduction of the dimer-dimer interactions as the dimers are 
compressed by the substrate, which destroys the HB ordering.
A CMC with non-ferromagnetic type ordering, such as the dimer HB phase,
arises only due to an effective multipole interaction between 
$n$-mers \cite{Trizac}. As $F_s$ increases
and the $n$-mers compress, the pole moment is reduced.    
We expect that similar types of dynamical behaviors
will arise for colloids on square substrates where similar 
spin type orderings are expected to occur \cite{Reichhardt,Trizac}. 

\begin{figure}
\includegraphics[width=3.5in]{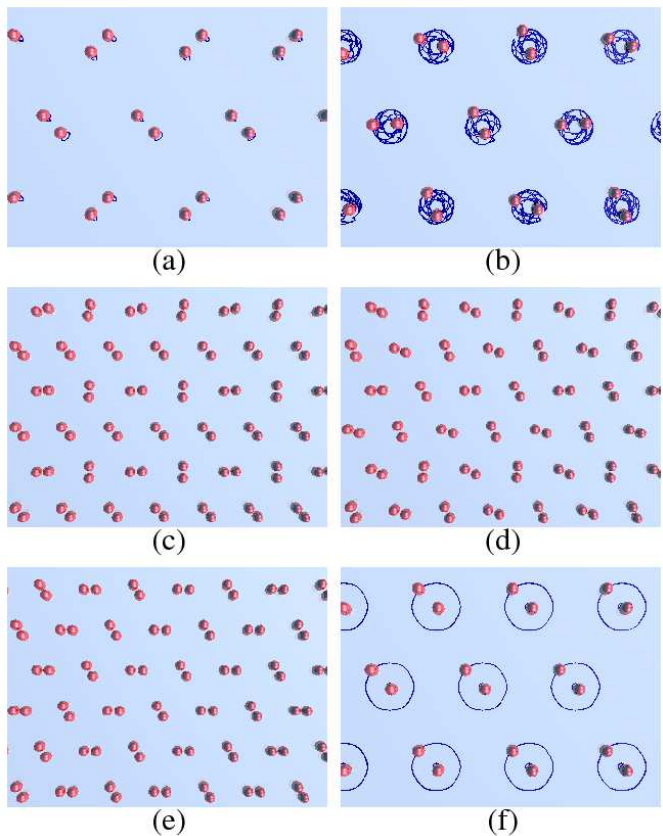}
\caption{ 
Colloid positions (balls) and trajectories (lines) 
in a portion of the sample over several ac
drive periods for a system with two colloids per substrate minima.
(a) The locked herringbone state at $F_s  = 3.0$ and $A_{ac} = 1.0.$ 
(b) The switching regime at $F_s = 3.0$ and $A_{ac} = 1.25$. 
(c) A snapshot of the colloid positions in the switching regime at the
initiation of a switch; (d) in the middle of the switch;
(e) upon completion of the switch.
(f) The continuously polarized phase at $F_s = 3.0$ and $A_{ac} = 2.0$.    
}
\end{figure}

In summary, we have shown that colloidal molecular crystals 
on periodic substrates 
exhibit a remarkable pattern of switching and polarization effects 
in the presence of a circular ac external drive.
We specifically examine a triangular substrate with
three and two colloids per trap which form ferromagnetic trimer or
herringbone dimer ground states.  The trimers show a switching regime in
which the partially polarized ferromagnetically ordered trimers
rapidly change their global orientation. 
The dimers exhibit structural switching
into different states as well as a martensitic switching between a 
herringbone and a ferromagnetically ordered state.
For strong substrates, both the dimers and the trimers enter a strongly or 
fully polarized state where the $n$-mers each develop a director that 
couples to the external drive.  This results in a continuous rotation of the
$n$-mers with the external drive 
which has similarities to liquid crystal dynamics.  
The ability to create colloidal molecular crystalline structures 
that can undergo rapid pattern switching or that have 
liquid crystal behaviors has exciting implications for 
creating a new class of photonic or phononic
drives, switches, displays, 
and new dynamical materials. Additionally, dynamical behaviors 
observed for the colloidal system
may be scaled down to systems 
such as Winger crystals on periodic substrates or 
crystalline cold atoms or molecules on optical
trap arrays.    
Realizing switching dynamics in these systems also has the potential 
to provide a basis for the development of
new types of nanodevices.

This work was carried out under the auspices of the 
NNSA of the 
U.S. DoE
at 
LANL
under Contract No.
DE-AC52-06NA25396.

\end{document}